\documentstyle[11pt,aasms,tighten,flushrt]{article}

\def \lesssim{\mathrel{<\kern-1.0em\lower0.9ex\hbox{$\sim$}}}
\def \gtrsim{\mathrel{>\kern-1.0em\lower0.9ex\hbox{$\sim$}}}

\begin{document}

\noindent  To appear in the proceedings of the conference,
  "Post-AGB Objects (proto-planetary nebulae) as a Phase of Stellar Evolution",
  Torun, Poland, July 5-7, 2000, eds. R. Szczerba, R. Tylenda, and S.K. Gorny.

\bigskip

\title{PLANETS AND AXISYMMETRIC MASS LOSS}

\author{
Noam Soker
\affil{
Department of Physics, University of Haifa at Oranim\\
Oranim, Tivon 36006, ISRAEL \\
soker@physics.technion.ac.il }}

$$
$$

\centerline {\bf ABSTRACT}
 Bipolar planetary nebulae (PNe), as well as extreme elliptical PNe
are formed through the influence of a stellar companion.  But half
of all PN progenitors are not influenced by any stellar companion,
and, as I show here, are expected to rotate very slowly on
reaching the upper asymptotic giant branch; hence they expect to
form spherical PNe, unless they are spun-up.  But since most PNe
are not spherical, I argue that $\sim 50 \%$ of AGB stars are
spun-up by planets, even planets having a mass as low as $0.01$
times the mass of Jupiter, so they form elliptical PNe.
The rotation by itself will not deform the AGB wind, but may
trigger another process that will lead to axisymmetric mass loss,
e.g., weak magnetic activity, as in the cool magnetic spots model.
This model also explains the transition from spherical to
axisymmetric mass loss on the upper AGB.  For such low
mass planets to substantially spin-up the stellar envelope, they
should enter the envelope when the star reaches the upper AGB.
This ``fine-tuning'' can be avoided if there are several planets on
average around each star, as is the case in the solar system, so
that one of them is engulfed when the star reaches the upper AGB.
Therefore I retain earlier predictions (Soker 1996) that on average
several planets are present around $\sim 50 \%$ of progenitors of PNe.


\section{ Introduction}

\subsection {Bipolar PNe: binary system progenitors (closed case)}

 I distinguish two main groups of nonspherical planetary nebulae (PNe):
bipolar and elliptical.
 Bipolar (also called ``bilobal'' and ``butterfly'') PNe
are defined (Schwarz, Corradi \& Stanghellini 1992) as
axially symmetric PNe having two lobes with an `equatorial'
waist between them; while elliptical PNe have a general elliptical shape.
Bipolar PNe amount to $\sim 10-15 \%$ of all PNe (Corradi \& Schwarz 1995).
The only physical property for which the difference between elliptical and
bipolar PNe is larger than the dispersion within each group is the
expansion velocity, which is much faster for bipolars.
 It is almost certain that bipolar PNe are formed from binary systems.
 In most cases the stellar companion stays outside the mass losing star
for the entire evolution, and in the rest it forms a common envelope
late in the evolution (Soker 1998a).
 I think that the observations, e.g., the similarity
of bipolar PNe to many symbiotic nebulae, and the theoretical
arguments, e.g., mechanism for blowing winds at
several$\times 100 ~{\rm km} ~{\rm s}^{-1}$, which support the
binary model for the formation of bipolar PNe, are extremely
strong (see summary in Table 1 by Soker 1998a).
 Some of the arguments can be found in Corradi (1995),
Corradi \& Schwarz (1995), Corradi {\it et al.} (1999, 2000),
Morris (1987, 1990), Mastrodemos \& Morris (1999),
Soker (1997, 1998a), Soker \& Rappaport (2000), and
Miranda {\it et al.} (2000).
 Several scenarios were proposed for the formation of
bipolar PNe from the evolution of single stars.
 I criticized these scenarios in different papers in the past,
and showed that they fail to explain basic facts, and/or they suffer
from unphysical arguments.
 I find the new paper by Matt {\it et al.} (2000) to fail on the same
main points.
 Some problems with these single star models are:
 (1) These models do not distinguish elliptical and bipolar
PNe, e.g., they can't account for the similarity of bipolar PNe and
symbiotic nebulae. 
 (2) The amount of angular momentum they require is much too large
for a single star, whether angular momentum directly to influence
the wind, or angular momentum required to amplify strong
magnetic activity. 
 (3) The single star models contradict the finding
that most of the 16 known PNe with central binary systems (Bond 2000)
possess an extreme elliptical shape, rather than a bipolar shape.
 Namely, if single stars can form bipolar PNe, how come AGB
stars which go through a common envelope phase, and hence rotate
much faster, form ``only'' elliptical PNe?
 These 16 PNe by themselves constitute very strong support for the
binary model for the formation of both bipolar and extreme
elliptical PNe (Bond \& Livio 1990; Soker 1997).

\subsection {Elliptical PNe: open questions}

 In addition to the $\sim 10-15 \%$ of all PNe which are bipolar,
and formed from binary systems, there are $\sim 20-30 \%$ elliptical
PNe formed from binary systems
(Yungelson, Tutukov \& Livio 1993; Han, Podsiadlowski \& Eggleton 1995;
Soker 1997).
 Most of these went through a common envelope phase and
have extreme structures (Bond \& Livio 1990; Bond 2000).
By extreme structure I refer to a large concentration of mass in
the equatorial plane, i.e., a torus, but there are no polar lobes,
hence the PN is not a bipolar PN.
 The AGB progenitors of the PNe were spun up by their
stellar companions to very high velocities, which led to
high mass concentration in the equatorial plane.
 The relevant axisymmetric mass loss mechanisms
for a common envelope evolution are summarized by Iben \& Livio (1993)
and Rasio \& Livio (1996), while axisymmetric mass loss mechanisms
for stellar companions outside the AGB envelope are listed
in Soker (1998a).

  But what about the rest of the axisymmetric PNe, 
which amount to $\sim 50 \%$ of all PNe?
 They were not spun up by a stellar companion, either because
the progenitor did not have a stellar companion, or the
companion is at a very large orbital separation.
 The open questions are therefore:
\newline
(1) What is the mechanism by which slowly rotating AGB stars can
blow axisymmetric winds?
Clearly centrifugal forces are unimportant.
\newline
(2) Why do many of the elliptical PNe have an outer spherical halo
while the inner region is elliptical?
 Put another way, why does the mass loss geometry change from 
spherical to axisymmetric only during the very end of the AGB
and/or during the early post-AGB phase?
(This question, and the answer given later, is the connection
of my paper to the title of the meeting.)
\newline
(3) Can a single star blow an axisymmetric wind during its final AGB
phase and/or post AGB phase, hence forming an elliptical PN?

\section {Slowly rotating AGB stars}
 I now give my answers to the questions raised in the previous
section.
 
{\bf  The axisymmetric mass loss mechanism},
was proposed several years ago (Soker 1998b) and was further
developed by Soker \& Clayton (1999) and Soker \& Harpaz (1999).
 It is assumed that a weak magnetic field forms cool stellar spots,
which facilitate the formation of dust closer to the stellar surface,
hence enhancing the mass loss rate there.
 If spots due to the dynamo activity are formed mainly near the
equatorial plane, then the degree of deviation from sphericity
increases.
 Based on a crude estimate I claimed (Soker 1998b)
that this mechanism operates for slowly rotating AGB stars,
having angular velocities of
$\omega{\mathrel{>\kern-1.0em\lower0.9ex\hbox{$\sim$}}}
10^{-4} \omega _{\rm Kep}$,
where $\omega_{\rm Kep}$ is the equatorial Keplerian angular velocity.
 I would like to stress that I do not propose a new mass loss mechanism.
I accept that pulsations coupled with radiation pressure on dust is
the mechanism for mass loss (e.g., Bowen 1988), and that the
luminosity, radius, and mass of the AGB star are the main
factors which determine the mass loss rate
(e.g., H\"ofner \& Dorfi 1997).
 I only suggest that cool magnetic spots facilitate the formation of dust,
and that their concentration near the equator causes the mass loss
geometry to deviate from sphericity (Soker 1998b; Soker \& Clayton 1999).
 It should also be noted that the required magnetic field is very weak,
has no direct dynamic effect, and is expected to form only a very weak
X-ray emission.
 Models based on strong magnetic fields (e.g., Matt {\it et al.} 2000)
are in contradiction with observations that AGB stars are weak
X-ray sources.

{\bf The transition to axisymmetric mass loss geometry},
in the cool magnetic spots model, is attributed to the shielding
of radiation by dust during the superwind phase (Soker 2000). 
 Soker (2000) proposed that dust which is formed very close to the
surface of a cool spot, practically at its surface, during a high
mass loss rate phase (superwind), has a large optical depth,
and it shields the region above it from the stellar radiation.
 As a result the temperature in the shaded region decreases rapidly
relative to the surrounding temperature.
 This leads to further dust formation in the shaded region.
 Without the formation of dust close to the surface of the spot and the
shielding, only large spots, with radii
$b_s{\mathrel{>\kern-1.0em\lower0.9ex\hbox{$\sim$}}} 0.3 R_\ast$,
allow enhanced dust formation (Frank 1995).
 This process is effective for small cool spots, but only when
mass loss rate is high, as in the superwind phase, hence optical
depth is large.
 Therefore, the equatorial enhanced mass loss rate occurs mainly
during the superwind phase at the end of the AGB.
 Another mechanism for the transition to axisymmetric mass loss,
which can operate in parallel to the the shielding of radiation,
is a more effective magnetic activity at the end of the AGB
(Soker \& Harpaz 1999).
In addition to the formation of elliptical PNe, the local
enhanced dust formation may lead to the formation of filaments,
loops, and arcs, as observed in many PNe.

{\bf Most single stars rotate too slowly} for the amplification
of even the weak magnetic field required by the cool magnetic
spots model.
 I argue that most are spun up by planets. This is the subject of
 the next section.

\section {The role of Planets}

 This section summarizes my recent paper (Soker 2001) in
which I examine the implications of the recently found
extrasolar planets on the planet-induced axisymmetric mass
loss model for the formation of elliptical PNe.
 I first show that single stars rotate very slowly as they
reach the upper AGB.
 I concentrate on stars with main sequence mass in the range
of $1.3 M_\odot < M_{\rm ms} < 2.4 M_\odot$.
 In this mass range the transition from slow main sequence rotators
to fast rotators occurs (e.g., Wolff \& Simon 1997), hence
these stars will clearly demonstrate the evolution of angular
momentum, while avoiding some uncertainties with lower mass
stars, e.g., the total mass they lose prior to the upper AGB
is not well known.
 I assume that the star rotates as a solid body (i.e., the angular
velocity is constant with radius inside the star) along
its entire evolution, and that the wind carries specific
angular momentum equal to that on the surface of the star.
 The average initial angular momentum on the main sequence is taken
from Wolf \& Simon (1997). 
  For the stars considered here, most of the mass loss occurs
on the AGB, when the mass of the core is $\sim 0.6 M_\odot$.
 Under these assumptions an analytical expression can be obtained for
the angular momentum (and angular velocity) on the upper AGB as
a function of the envelope mass retained by the star as it loses
mass (Soker 2001).
  As an example I present the result for a single star evolving
on the upper AGB, and which had a mass of $M_{\rm ms}=1.8 M_\odot$
on the main sequence. 
 The figure shows the evolution of the angular velocity (solid line),
in units of $\omega_{\rm Kep}$, and the angular momentum (dashed line),
in units of the orbital angular momentum of Jupiter $J_J$, as a
function of the envelope mass left in the envelope.
 We note the fast decrease of the angular velocity as envelope
mass decreases due to mass loss.

 In the cool magnetic spots model the role of the rotation is
mainly to shape the magnetic field into an axisymmetric configuration
(on average), and it may operate efficiently even for an envelope
rotating as slowly as $\omega \sim 10^{-4} \omega_{\rm Kep}$
(Soker \& Harpaz 1999).
 From the figure we see that single stars will not possess the required
angular velocity when the envelope mass decreases below $\sim 0.3 M_\odot$.
 However, very low mass planets, down to $\sim 0.01 M_J$, where $M_J$
is Jupiter's mass, are sufficient, if they enter the
AGB envelope at late stages.
 For example, a planet of mass $0.01 M_J$ at an orbital separation
of $2 ~{\rm AU}$ has an angular momentum about equal to
that of an AGB star with envelope mass of $M_{\rm env} =0.4 M_\odot$
which had a main sequence mass of $M_{\rm ms} =1.8 M_\odot$.
 If such a planet enters the envelope when $M_{\rm env}=0.2 M_\odot$,
for example, it will increase the AGB envelope angular momentum
by a factor of $\sim 30$.  
 Taking the solar evolution to the AGB (see Soker 2001) I find that
when the envelope mass becomes $0.15 M_\odot$, the angular momentum
of the AGB sun is $\sim 10^{-4} J_J$, or $\sim 0.1$ the angular
momentum of Earth.
  By that time the orbital separation will be $1.33 ~{\rm AU}$
(or $290 R_\odot$).
  If the sun at this stage goes through a helium shell flash, so that
the radius increases, say, to $\sim 1.3 ~{\rm AU}$ then another $\sim 10 \%$
increase during the maximum radius in the pulsation cycles may reach
the location of Earth, causing the Earth to spiral inside the solar
envelope.
 A detailed analysis of the evolution of the Earth-sun system, until the
sun leaves the AGB, for different assumptions and models,
is given by Rybicki \& Denis (2000).
   As a result of the deposition of the Earth's orbital momentum,
the solar envelope will rotate $\sim 10$ times faster,
or at $\omega \simeq 10^{-4} \omega_{\rm Kep}$.
 If this occurs indeed in about 7 billion years, then the Earth may be
responsible for the PN of the sun being elliptical rather
than spherical.
 However, it is not clear that the sun will engulf the Earth,
or that it will form a PN at all (Rybicki \& Denis 2000).

 For a high probability that a planet will enter the AGB
envelope at late stages, i.e., for it to occur in many stars,
two things should happen.
First, on average there should be several planets around each
star (as is the case in the solar system), and second,
there should be a fast and significant increase of the stellar radius
on the upper AGB.
 Numerical simulations of AGB stars show that after thermal pulses
(helium shell flashes) on the upper AGB, the envelope increases
by $\sim 20-30 \%$.
 This is in addition to the increase in the average AGB stellar radius
as the core mass increases.
 So the second condition is fulfilled for upper AGB stars.
 The first condition is a requirement, hence a {\it prediction},
of the planet-induced axisymmetric mass loss model for the formation
of elliptical PNe.
 The new addition of the present paper is the relaxation of the
minimum mass demand on planets from $\sim 1 M_J$
(Soker 1996) to $\sim 0.01 M_J$.
The motivations for reducing the lower mass limit are
the new finding that only $\sim 5 \%$ of sun-like stars have
Jupiter-like planets around them, and a new model for
axisymmetric mass loss, the cool magnetic spots model, which
was constructed to work for very slowly rotating AGB stars,
as discussed above.

 Finally, note that many of the known sun-like
stars that have planets around them will not form PNe at all.
This is because their orbiting planet will spin-up the envelope
and deposit energy already on the stellar red giant branch (RGB),
hence mass loss on the RGB is expected to be high, and most of
the stellar envelope will be lost already on the RGB.
 No observable nebula will be formed.
  So, while in most cases planet companions will lead to the
formation of an elliptical rather than a spherical PN, in some cases
Jupiter-like planets in close orbits around low mass stars
will prevent the stars from forming a PN.
    
\acknowledgments
This research was supported in part by grants from the 
Israel Science Foundation and the US-Israel Binational Science Foundation.

{\bf Figure 1:}
\newline
 Evolution of the angular momentum and angular velocity
as a function of the mass left in the AGB stellar envelope
of a star which had a mass of $1.8 M_\odot$ on the main sequence
(for details see Soker 2001).
Angular momentum is in units of Jupiter's orbital angular
momentum, and angular velocity in units of the Keplerian angular
velocity on the stellar equator.

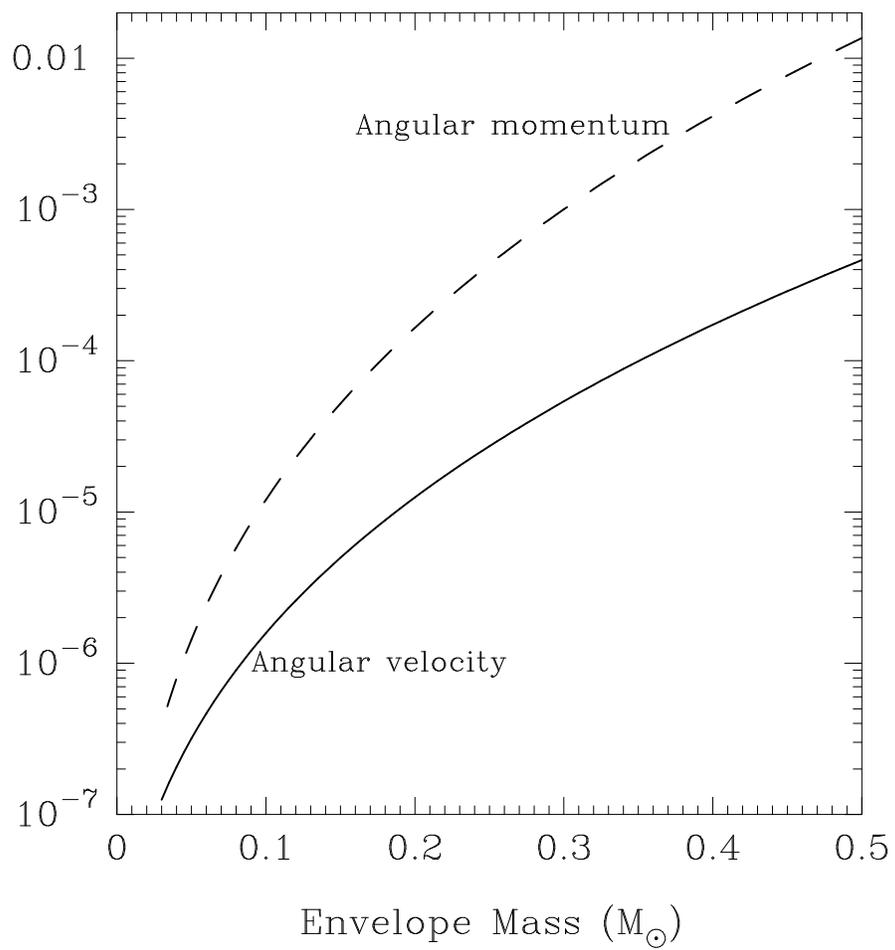
\end{document}